\documentclass[prd,twocolumn,showpacs,preprintnumbers,floats,floatfix,nofootinbib]{revtex4}

\usepackage{graphicx}%
\usepackage{dcolumn}
\usepackage{amsmath}
\begin{document}

\title{The low dimensional dynamical system approach in General
Relativity: an example}

\author{H. P. de Oliveira}
\email{oliveira@dft.if.uerj.br}
\author{E. L. Rodrigues}
\email{elrodrigues@uerj.br}
\affiliation{{\it Instituto de F\'{\i}sica, Universidade do Estado do Rio de Janeiro }\\
{\it R. S\~ao\ Francisco Xavier, 524, CEP 20550-013\\
Rio de Janeiro, RJ, Brazil}}

\author{I. Dami\~ao Soares}
\email{ivano@cbpf.br}
\affiliation{{Centro Brasileiro de Pesquisas F\'{\i}sicas/MCT}
\\{\it R. Dr. Xavier Sigaud, 150, CEP 22290-180}
\\{\it Rio de Janeiro, RJ, Brazil}}

\author{E. V. Tonini}
\email{tonini@cefetes.br}
\affiliation{{Centro Federal de Educa\c c\~ao Tecnol\'ogica do Esp\'\i rito Santo}\\
{Avenida Vit\'oria, 1729, CEP 29040-780}
\\{Vit\'oria, ES, Brazil}}

\date{\today}

\begin{abstract}
\noindent In this paper we explore one of the most important features of the
Galerkin method, which is to achieve high accuracy with a
relatively modest computational effort, in the dynamics of
Robinson-Trautman spacetimes.
\end{abstract}

\maketitle

\section{Introduction}%

\par The General Theory of Relativity (GR) constitutes one of
the major scientific achievements of the past hundred years. Its
theoretical framework establishes that the gravitational field and
the geometry of the spacetime are the same physical entity and that
the equations ruling the gravitational dynamics are field equations
of the spacetime geometry. GR has so far passed extremely well all
experimental tests\cite{exp_tests}, in opposition to other
relativistic extensions as scalar-tensor theories. Basically, GR
establishes a system of ten nonlinear, coupled partial differential
equations that governs the dynamics of the gravitational field
represented by the symmetric second order metric tensor
$g_{\alpha\beta}$, $\alpha,\beta=0,1,2,3$. The nonlinear nature of
the field equations is the main obstacle to obtain exact solutions
of the field equations, unless we make appeal to symmetries in the
solutions, or alternatively the use of perturbative approaches are
adopted. Therefore, in order to study the dynamics of the
gravitational field in more general situations the use of numerical
techniques seems to be the only possible strategy to circumvent the
difficulty posed by the nonlinearities of the field equations. In
this context numerical relativity has become a very fertile and at
the same time challenging field of research as recently considered
in several interesting reviews\cite{winicour}. There the improvement
of specific numerical techniques adapted to relativistic problems,
along with the increase of computational resources figure as
promising factors for the advance of numerical relativity.
Nonetheless, despite all the progress made so far the complete
understanding of important problems in relativistic astrophysics
such as non-spherical collapse and nonlinear regimes of emission of
gravitational waves is still not complete.
\par A promising approach to treat numerically nonlinear problems is
provided by the so-called spectral methods\cite{spectral}. The
spectral methods adopt a distinct strategy if compared with the
finite difference scheme. For instance, considering a function
$u(t,x)$ satisfying a given one dimensional partial differential
equation, it will be approximate as a series of the type $u_a(t,x)
= \sum_{k=0}^N\,a_k(t) \psi_k(t)$, where the basis or trial
functions $\psi_k(x)$ are known analytical polynomials such as
Fourier, Legendre, Chebyshev, etc. In general, by increasing the
truncation order $N$, $u_a(t,x)$ approaches of the exact solution
of the problem in the mean. There are distinct types of spectral
methods among which we list the Galerkin
method\cite{fletcher,holmes}, the collocation method\cite{boyd}
and the Tau-method. These methods have an attractive feature which
is to transform any partial differential equation into a finite set
of coupled ordinary differential equations, or simply a dynamical system
whose dimension is dictated by the truncation order $N$. Another
important and robust feature is the high accuracy achieved with a
small truncation order, what corresponds to use moderate or low
computational resources.
\par In this paper our objective is to present a consistent low
dimensional dynamical system approach provided by the Galerkin
method applied to the problem of non-spherical collapse with
the emission of gravitational waves, in the realm of Robinson-Trautman
geometries\cite{rt}. In this way, the paper is
divided as follows. The basic equations that govern the dynamics
of Robinson-Trautman spacetimes are presented in Section 2. In
Section 3 we discuss and implement efficiently the Galerkin
method. In Section 4 numerical experiments are performed in order to
test the efficiency and convergence of the method. Section 5 is
devoted to the conclusions and final remarks.
%
%
\section{The dynamics of Robinson-Trautman spacetimes: the nonlinear differential equation}

${ }$\par Robinson-Trautman (RT) spacetimes are the simplest
axisymmetric spacetimes that can be viewed as the exterior
geometry of an isolated source emitting gravitational
waves\cite{rt,nunti} or the exterior of a distorted black hole.
In a suitable coordinate system the line element for RT
geometries is expressed as

\begin{eqnarray}
ds^2&=&\left(\lambda(u,\theta) - \frac{2 m_{0}}{r} + 2 r
\frac{\dot{K}(u,\theta)}{K(u,\theta)}\right) d u^2 + 2 du dr -
\nonumber \\
& &  r^{2}K^{2}(u,\theta)(d \theta^{2}+\sin^{2}\theta d
\varphi^{2}), \label{eq1}
\end{eqnarray}

\noindent where $m_{0}$ is a constant related to the total mass of
the system and dot means derivative with respect to $u$. According
to Einstein field equations the function $\lambda(u,\theta)$
is related to $K=K(u,\theta)$ by the constraint equation

\begin{equation}
\label{eq2} \lambda(u,\theta)=\frac{1}{K^2}-\frac{K_{\theta
\theta}}{K^3}+\frac{K_{\theta}^{2}}{K^4}-\frac{K_{\theta}}{K^3}\cot
\theta,
\end{equation}

\noindent where the subscript $\theta$ denotes derivative with
respect to the angle $\theta$. The remaining vacuum Einstein equations
impose that the function $K(u,\theta)$ satisfies the evolution equation

\begin{equation}
\label{eq3} -6 m_{0}\frac{\dot{K}}{K}+\frac{(\lambda_{\theta} \sin
\theta)_{\theta}}{2 K^2 \sin \theta}=0,
\end{equation}

\noindent which henceforth will be denoted as the Robinson-Trautman
(RT) equation. This equation governs the evolution of the
gravitational field, in other words, it allows to evolve the
initial data $K_{0}(\theta )=K(u=u_{0},\theta)$ prescribed on a
given null surface $u=u_0=constant$ (except in the case $m_0=0$).
Despite its relatively simple form no general exact non-stationary
solution of RT equation is known. However by restricting to stationary
configurations two important solutions are known. The first is the
Schwarzschild solution that represents the gravitational field a
black hole characterized by $K=K_0=constant$, which implies
$\lambda=K_0^{-2}$ with a corresponding Schwarzschild mass $M=m_{0}K_{0}^3$.
The second solution represents a boosted black
hole\cite{bondi} along the axis of symmetry such that

\begin{equation}
\label{eq4} K(\theta )=\frac{1}{\cosh \gamma +\cos \theta \sinh
\gamma},
\end{equation}

\noindent where $\gamma$ is a parameter associated with the velocity
of the boost, $v=\tanh \gamma$. The constraint equation (\ref{eq2}) in this case
yields $\lambda = 1$. 

\section{The spectral method approach}

It will be useful to express $K(u,\theta )$ as

\begin{equation}
\label{eq5} K(u,x)=A_{0}e^{\frac {1}{2}Q(u,x)},
\end{equation}

\noindent where $A_0$ is a constant, and for convenience we have
introduced the variable $x=\cos \theta$ with
$-1\leq x \leq 1$. The first step to apply the Galerkin method is
to express $Q(u,x)$ as a power series with respect to a suitable
set of basis functions chosen as the Legendre polynomials due to
the symmetry of the problem and the required regularity of $K(u,x)$
with respect to $x$. In this way an approximated expression for $Q(u,x)$
is established and given by

\begin{equation}
\label{eq6} Q_{a}(u,x)=\sum_{k=0}^{N}{b_{k}(u)P_{k}(x)},
\end{equation}

\noindent where $N$ is the truncation order, $b_{k}(u)$ and
$P_{k}(x)$ are the modal coefficients and the Legendre polynomials
of kth order. This basis functions define an abstract projection
space for which an internal product can be defined, in this case,
as

\begin{equation}
\label{eq7} \left\langle P_{j}(x),P_{k}(x)\right\rangle
=\int_{-1}^{+1}{P_{j}(x)P_{k}(x) dx = \frac{2\delta _{kj}}{2k+1}}.
\end{equation}

\noindent The next step is to substitute the Galerkin
decomposition for $Q(u,x)$ into Eq. (\ref{eq2}) to obtain an
approximated expression for $\lambda(u,x)$, or

\begin{equation}
\label{eq8} \lambda
_{a}(u,x)=\frac{e^{-Q_{a}(u,x)}}{A_{0}^{2}}\left(1+\sum_{k=0}^{N}{\frac{1}{2}k(k+1)b_{k}(u)P_{k}(x)}\right).
\end{equation}

The so called residual equation follows after introducing the
approximated expressions for $Q(u,x)$ and $\lambda(u,x)$ given
above into the RT equation (\ref{eq3}). After some direct
calculation we obtain schematically

\begin{eqnarray}
{\rm{Res}}(u,x)=\sum_{k=0}^{N}{b_{k}(u)P_{k}(x)}-\frac{{\rm{e}}^{-Q_{a}(u,x)}}{6
m_0 A_{0}^{2}}\left[(1-x^2)\lambda _{a}^{\prime}\right]^{\prime},
\label{eq9}
\end{eqnarray}

\noindent where prime stands for derivative with respect to $x$.
In fact the residual equation can be a good measure of the error
of truncation and a valuable test\cite{finlayson} for the
convergence of the Galerkin decomposition. According to the
traditional Galerkin method the projection of the residual
equation into each basis function $P_{n}(x)$, $n=0, 1,..., N$ is
assumed to be zero in order to minimizing the error of truncation,
or $\left<{\rm{Res}}(u,x),P_{n}(x)\right> = 0$, rendering the
following dynamical system for the modal coefficients

\begin{eqnarray}
\label{eq10} \dot{b}_{n}(u)&=& \frac{(2 n+1)}{12 m_0
A_0^2}\,\int_{-1}^1\,\exp\left(-\sum_{k=0}^{N}{b_{k}(u)P_{k}(x)}\right)\,\times
\nonumber \\
& & [(1-x^2)\,\lambda_a^{\prime}]^{\prime} P_n(x)\,d x,
\end{eqnarray}

\noindent with $n=0, 1, ..., N$.

Here some remarks are necessary. In order to express the rhs of
the above equation in terms of the modal coefficients, it is
necessary to integrate a large expression involving an exponential
of a series of Legendre functions, the size of which depends on
the truncation order $N$. This integration must be necessarily
done symbolically (in Maple or Mathematica), but unless $N \leq 2$
- a very low truncation order -  such an integration cannot be
performed. Thus, we can adopt two distinct strategies based on
further approximation schemes:

\begin{itemize}

\item The first is to assume

\begin{equation}
\label{eq11} e^{-Q_a}=1 - Q_a + \frac{Q_a^{2}}{2!}+ ... \cong
\sum_{j=0}^{J}{\frac{(-Q_a)^{j}}{j!}},
\end{equation}

\noindent which renders a polynomial expression for the
exponential. Introducing this expression into Eq (\ref{eq9}) the
integration can be done promptly without much cost of CPU time for
reasonable choices of $N$ and $J$. As a matter of fact, this
additional approximation can be acceptable for not large values of
$|Q|$ as we are going to see later.

\item The second strategy consists, on the hand,  in applying the
collocation method\cite{boyd} to express ${\rm{e}}^{-Q_{a}(u,x)}$
as a series of another basis functions, or

\begin{equation}
\label{eq12} {\rm{e}}^{-Q_{a}(u,x)} \cong
\sum_{k=0}^{M}{q_{k}(u)T_{k}(x)},
\end{equation}

\noindent where $T_{k}(x)$ are the Chebyshev polynomials, and
$q_k(u)$ the new modal coefficients given as functions of the
original coefficients $b_j(u)$. These relations are fixed by the
fact that at the collocation points $x_{j} = \cos\left(\pi
j/M\right)$, $j = 0,1,...,M$ the above decomposition agrees with
the exact expression, or

\[
{\rm{e}}^{-Q_{a}(u,x_j)} = \sum_{k=0}^{M}{q_{k}(u)T_{k}(x_j)},
\]

\noindent and together with the discrete orthogonality relation
$\delta_{jk} = \frac{2}{M\bar{c}_k}
\sum_{n=0}^M\,\frac{1}{\bar{c}_n}T_j(x_n)T_k(x_n)$, where
$\bar{c}_0 = \bar{c}_M = 2$ and $\bar{c}_k = 1$, for $1 \leq k
\leq M-1$. As a consequence each modal coefficient $q_k$ can be
expressed in terms of $b_k$. In both schemes a $N+1$ dimensional
dynamical system for the modal coefficients is obtained after
integrating the $rhs$ of Eq. (\ref{eq10}).

\end{itemize}

\section{Numerical experiments}

Our first numerical task will be to study the accuracy and
convergence of the approximative schemes outlined at the end of
the last Section. To this end we consider the exact expressions
for some initial data $K_0(x)=K(u_0,x)$ identified as an oblate
spheroid and a black hole in motion given, respectively, by

\begin{eqnarray}
\label{eq13}
K_0(x) = \frac{k_0 \sqrt{1-e^2}}{\sqrt{1-(1-x^2)e^2}} \nonumber
\\
K_0(x) = \frac{1}{\cosh \gamma + x \sinh \gamma},
\end{eqnarray}

\noindent where $k_0$ and $\gamma$ are constants and $e$ is the
eccentricity. As a further assumption the eccentricity can depend
on $x$ as $e=e_0+e_j x^j$, with $e_0$, $e_j$ being parameters, in
order to encompass the usual oblate spheroid ($e_j=0$ and $0 < e_0
< 1)$, and a deformed oblate spheroid ($e_0>e_j \neq 0$). It is
worth of calling attention that no dynamics is present if the
second data family is chosen, since it represents an exact
stationary solution of the RT equation.

\begin{figure}[ht]
\begin{center}
\rotatebox{270}{\includegraphics*[height=7cm,width=5.5cm]{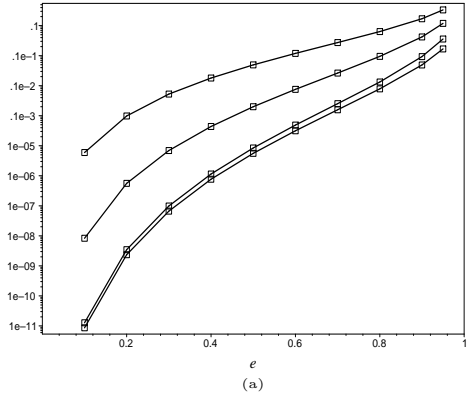}}
\centerline{\tiny{(a)}} 
\rotatebox{270}{\includegraphics*[height=7cm,width=5.5cm]{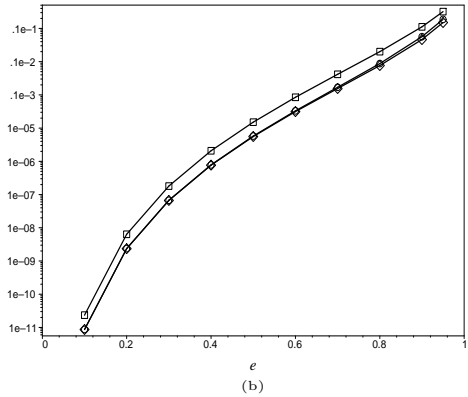}}
\centerline{\tiny{(b)}} \caption{$L_2$ error for the oblate
spheroid ($e_j=0$) in function of the eccentricity $e$ for
$J=1,2,3,4$ (upper to lower curves). Similar plots are constructed
in (b) considering the second scheme of approximation, for which
the we have selected $M=6,8$ and 10 collocation points starting
from the upper curve (boxes). Note that the last two curves almost
coincide indicating a rapid convergence of the method, whereas in
(a) the convergence is evident only if $J \geq 3$. In all cases we
have set $N=7$.}
\end{center}
\end{figure}

According to the previous discussion we choose the truncation
order $N$ from which the decomposition
$Q_{a}(u_0,x)=\sum_{k=0}^{N}{b_{k}(u_0)P_{k}(x)}$ is established.
The values of each modal coefficient $b_k(u_0)$ will be determined
using the standard Galerkin projection procedure starting from
$K_0(x) \simeq K_a(x) = A_0 \exp(1/2
\sum_{k=0}^{N}{b_{k}(u)P_{k}(x)})$ that yields

\begin{equation}
\label{eq14} b_k(u_0) = \frac{\left<2 \ln (A_0^{-1}
K_0(x),P_k(x)\right>}{\left<P_k(x),P_k(x)\right>}.
\end{equation}

\noindent Henceforth, we set $A_0=k_0=1$ without loss of
generality. Once the modal coefficients are determined $\exp(1/2
Q_a)$ will be further approximated according to the schemes
presented in the last Section: the first is to expand the
exponential as a power series of $-Q/2$ in the same way as shown
in Eq. (\ref{eq11}); the second is to expand the exponential using
the collocation method. In order to provide a quantitative measure
of the error between these approximations and the \textit{exact}
expressions for the corresponding initial data, the $L_2$ error
defined by

\begin{equation}
L_2 = \sqrt{\frac{1}{2}\int_{-1}^1\,[K_0(x)-K_a(x)]^2 dx}
\label{eq15}
\end{equation}

\noindent will be evaluated in each case.

\begin{figure}[ht]
\begin{center}
\rotatebox{270}{\includegraphics*[height=7cm,width=5.5cm]{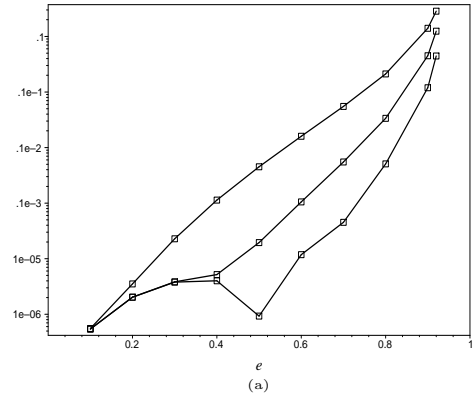}}
\centerline{\tiny{(a)}}
\rotatebox{270}{\includegraphics*[height=7cm,width=5.5cm]{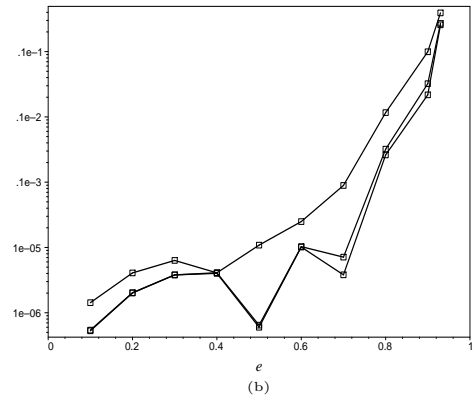}}
\centerline{\tiny{(b)}}
\caption{$L_2$ error for the deformed oblate spheroid ($e_2=0.07$)
in function of the eccentricity $e_0$ for (a) $J=2,3,4$ (upper to
lower curves). Similar plots are constructed in (b) considering
the second scheme of approximation, for $M=6,8$ and 10 collocation
points starting from the upper curve. Note that as shown in Fig.
1(b) the last two curves almost coincide indicating a rapid
convergence of the method. In all cases we have set $N=7$.}
\end{center}
\end{figure}

Let us consider the oblate spheroid with $e_j=0$ and $0 < e <1$. We
then fix the truncation order to $N=7$ and evaluate the error $L_2$
between the exact and approximate expressions for the oblate
spheroid taking into account both additional schemes as previously
described. In the first we choose $J=1,2,3$ and 4 present in the
expansion $\exp(Q_a/2) \simeq \sum_{k=0}^{J}\,2^{-k}Q_a^k/k!$,
whereas for the second scheme the expansion of $\exp(Q_a/2)$ is
constructed with $M=6,8$ and 10 collocation points. The respective
$L_2$ errors evaluated for successive values of the eccentricity $e$
are shown in Fig. 1. Two important aspect are worth to be mentioned.
As expected small eccentricities produce very small errors even for
the quite modest choice $J=1$, since these cases correspond to tiny
departures from the spherically symmetric configuration described
exactly by the Galerkin decomposition as $b_0(u_0)=$ constant,
$b_k(u_0) = 0$ for $k \neq 0$. High eccentricities, otherwise,
represent more severe deviations from the spherical configuration,
producing as a consequence an increase of the error. The second
aspect concerns to the fast convergence achieved using both
approaches as suggested by the plots of Figs. 1(a) and 1(b).
Accordingly, in the first case the convergence is clear if $J \geq
3$, and it is necessary that the number of collocation points is
greater than, or $N \geq 8$\footnote{The convergence depends on the
truncation order, but we have noticed that for a fixed truncation
order $N$ the most efficient decomposition is obtained if $M=N+1$
collocation points is selected. If $M > N+1$ there is not relevant
improvement in the error as exemplified in Fig. 1(b).}. Despite the
acceptable errors attained with a modest truncation order ($N=7$) in
both cases, the use of the second scheme appears to be superior
because the less algebraic effort necessary to achieve the same
results of the first case with $J=4$.

In Fig. 2 we have evaluated the $L_2$ error for the (deformed)
oblate spheroid with variable eccentricity $e=e_0+e_2x^2$, in
which we have set $e_2 = 0.07$ and again $0 < e < 1$. As it can be
seem from the plots it is clear that the scheme using the
collocation points is more efficient, in the sense of the need of
smaller computational effort, and accurate than the first scheme.
The initial data representing a black hole in motion is considered
in Fig. 3, where $J=4$ and $M=8$ collocation points which confirms
once more the previous conclusion.



\begin{figure}[ht]
\begin{center}
\rotatebox{270}{\includegraphics*[height=7cm,width=5.5cm]{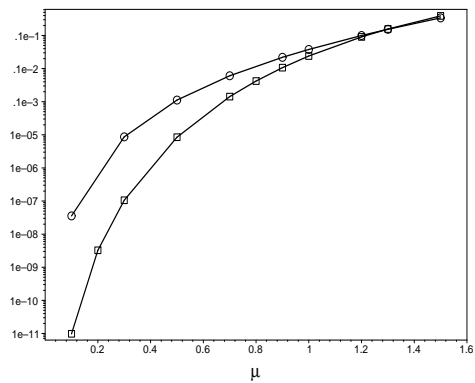}}
\caption{$L_2$ error for the initial data identified as the
boosted black hole (\ref{eq13}) if the truncation order is $N=7$,
and taking into account the collocation scheme ($M=8$, lower
curve) and the first scheme with $J=4$ (upper curve) for the first
scheme.}
\end{center}
\end{figure}

In the second group of numerical experiments we shall verify the
evolution of the error associated to the constraint equation
(\ref{eq2}). The procedure consists into rewrite Eq. (\ref{eq2})
in the following way

\begin{equation}
\label{eq16} \lambda K^2+\left[(1-x^2)
\frac{K^\prime}{K}\right]^\prime-1 = 0.
\end{equation}

\noindent Here $\lambda(u,x)$ is given by Eq. (\ref{eq8}), but
$K(u,x)$ will be expressed according to the schemes of
approximation as outlined by Eqs. (\ref{eq11}) and (\ref{eq12}).
For instance, by assuming the expansion provided by the later
expression we have

\begin{equation}
K(u,x) = A_0^2\,{\rm{e}}^Q \simeq A_0^2
(\left(\sum_{k=0}^{M}{q_{k}(u)T_{k}(x)}\right)^{-1}, \nonumber
\end{equation}

\noindent and, as a consequence, the constraint equation now reads

\begin{eqnarray}
& & {\rm{CE}}(u,x) = \lambda_a
A_0^2\,\left(\sum_{k=0}^{M}{q_{k}(u)T_{k}(x)}\right)^{-1}
\nonumber \\
& & - \frac{1}{2}\left[(1-x^2)
\frac{\sum_{k=0}^{M}{q_{k}(u)T^\prime_{k}(x)}}{\sum_{k=0}^{M}{q_{k}(u)T_{k}(x)}}\right]^\prime
- 1 = 0.
\end{eqnarray}

\begin{figure}[h]
\begin{center}
\rotatebox{270}{\includegraphics*[height=7cm,width=5.5cm]{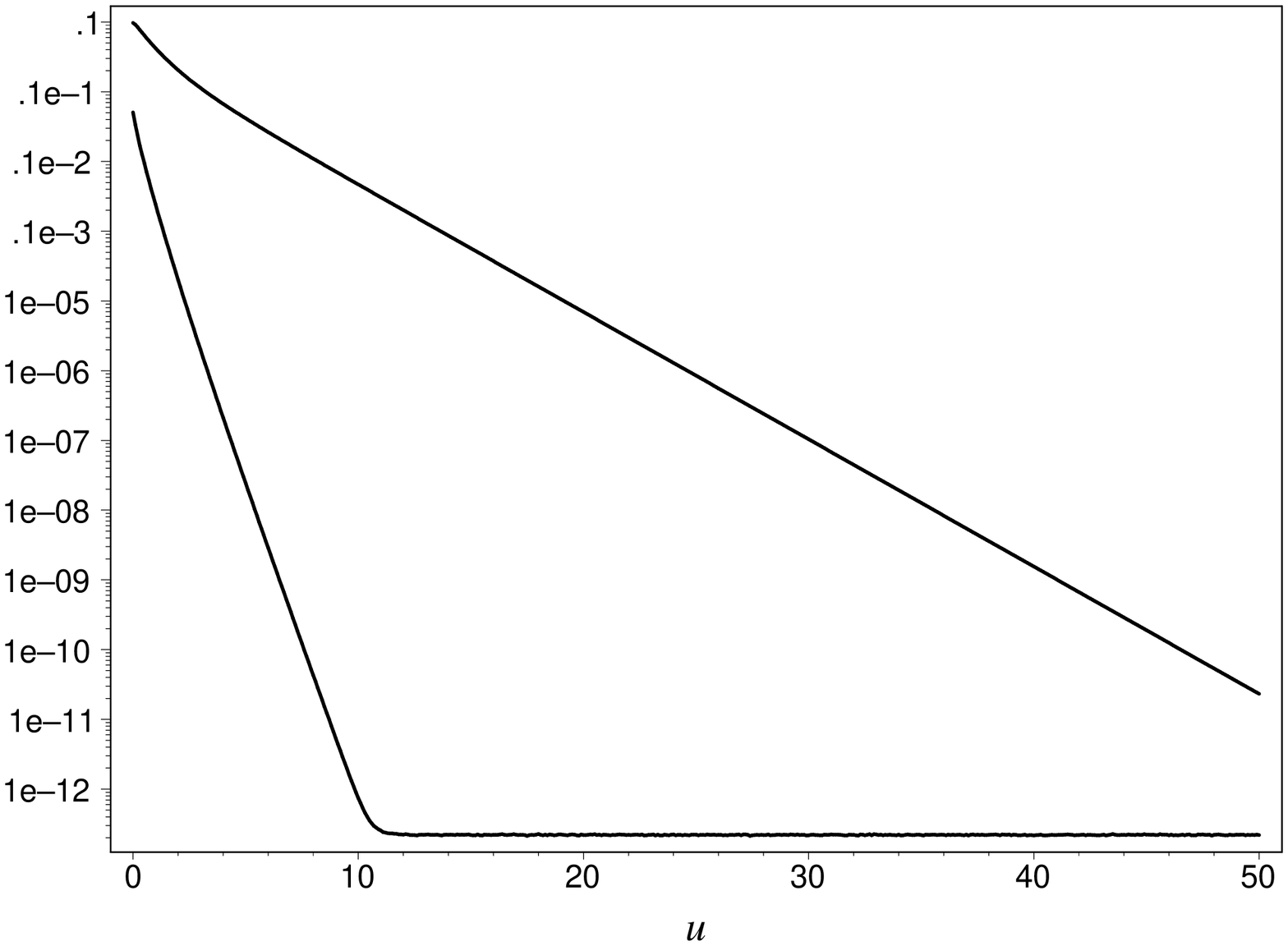}}
\centerline{\tiny{(a)}}
\rotatebox{270}{\includegraphics*[height=7cm,width=5.5cm]{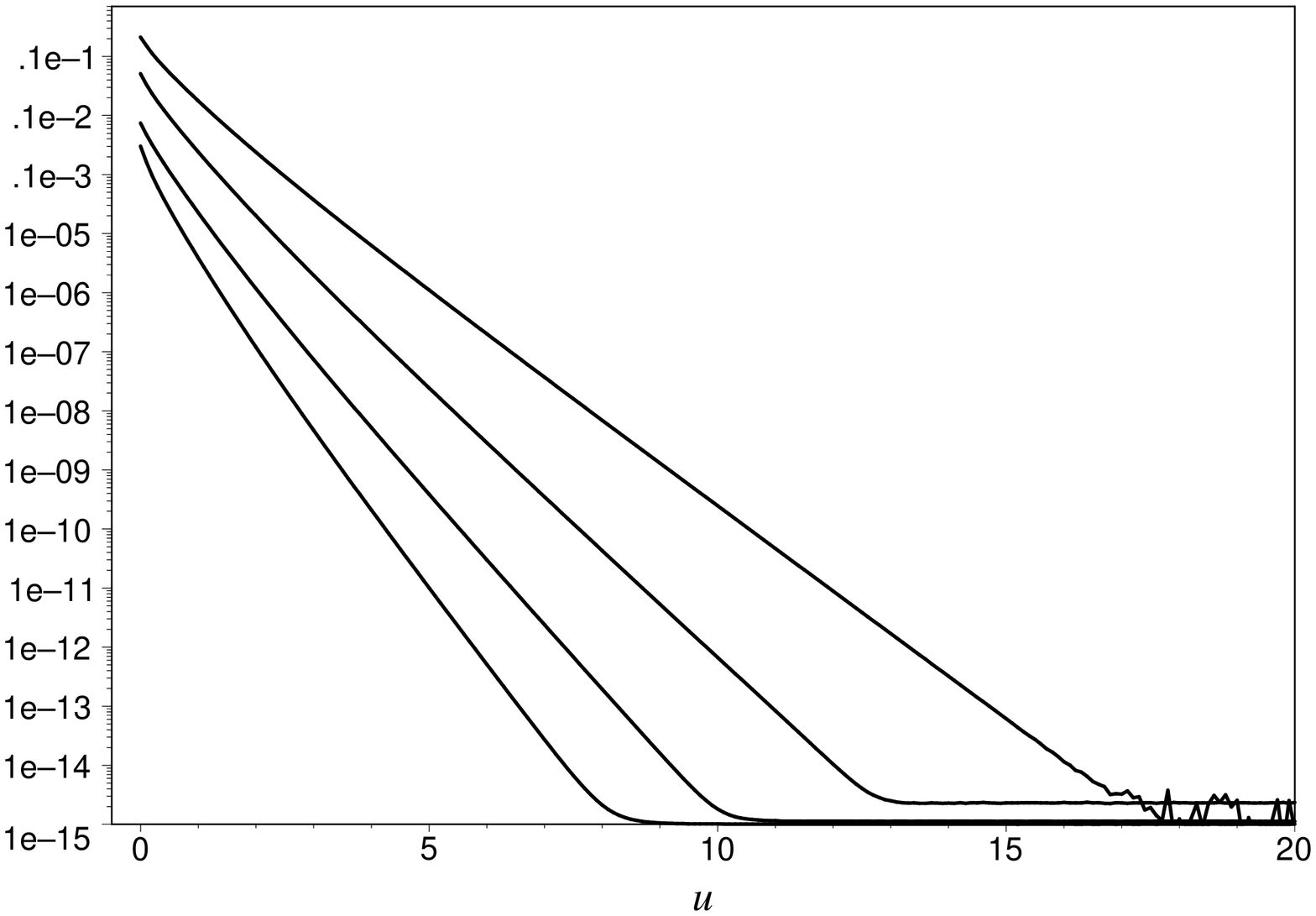}}
\centerline{\tiny{(b)}}
\caption{(a) Evolution of the $L_2$ error of the constraint equation
using the first and the second schemes of approximation (upper and
lower curves, respectively). Here $N=7$, $J=2$ (see Eq.
(\ref{eq11})) and $M=8$ collocation points, where it is clear the
rapid convergence of the collocation expansion (\ref{eq12}). In (b)
we present the influence of increasing the truncation orders
$N=5,7,9,11$ with respective $N+1$ collocation points. The initial
conditions correspond to the oblate spheroid with eccentricity
$e_0=0.8$.}
\end{center}
\end{figure}

\noindent A similar expression is obtained if the approximation
expansion given by $(\ref{eq11})$ is assumed. In any case the
resulting form of the constraint equation is evaluated at each
instant after the integration of the dynamical equation for the
modal coefficients. As a matter of fact, we are interested in the
rms error of the constraint or

\begin{equation}
\sqrt{\frac{1}{2}\int_{-1}^{1}\,{\rm{CE}}(u,x)^2 dx}
\end{equation}


\noindent Fig. 4 shows the evolution of this error if the initial
data is the oblate spheroid ($e_0=0.8$, $e_j=0$) for both schemes
of approximation.


The residual equation can provide another possible way of
verifying the convergence and accuracy of the numerical method as
the truncation order is increased. In fact, as demonstrated by ??,
the rms error of residual equation is identified as the upper
bound of the rms error between the approximate and exact
solutions. Then, in our last numerical experiment the evolution of
the error

\begin{equation}
\nonumber \sqrt{\frac{1}{2}\,\int_{-1}^1 {\rm{Res}}(u,x)^2 dx},
\end{equation}

\noindent where ${\rm{Res}}(u,x)$ is the residual equation (cf.
Eq. (\ref{eq9})) considering only the second scheme of
approximation (Eq. (\ref{eq12})) based on the collocation method.
In Fig. 5 the convergence is illustrated by graphs corresponding
to several truncation orders, namely, $N=5,7,9,11$, and
respectively $M+1$ collocation points. As it can be seem the error
decreases rapidly until reaching to the value considered zero up
to our numerical precision, and the greater is the truncation
order less time is necessary to attain this value.

\begin{figure}[ht]
\begin{center}
\rotatebox{270}{\includegraphics*[height=8cm,width=6cm]{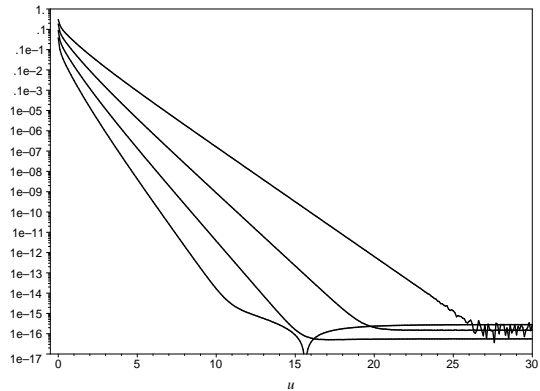}}
\caption{Evolution of the residual error the truncation orders
$N=5,7,9,11$ (upper to lower curves) and respective $N+1$
collocation points. The initial conditions correspond to the oblate
spheroid with eccentricity $e_0=0.8$.}
\end{center}
\end{figure}

\section{Final remarks}

The low dynamical system approach provided by the Galerkin method
constitutes a valuable tool to deal with nonlinear problems as
demonstrated not only here, but also in problem of turbulence of
fluids\cite{holmes}. Two aspects are worth of mentioning:

\begin{itemize}
\item The transformation of, say, a partial differential equation
to a finite set of ordinary differential equations. In general the
greater is the set of these equations more accurate is the
approximate solution.

\item A reasonable accuracy can in fact be obtained with a
relatively low dimensional dynamical system.
\end{itemize}

\noindent As a matter of fact we have applied the Gakerkin method
to a problem of gravitation, namely, the dynamics of the
Robinson-Trautman spacetimes that represents the simplest
axisymmetric geometries endowed with gravitational waves. As we
have described one of the crucial steps to implement the method is
to project the residual equation into each basis function, which
is necessary to obtain the dynamical system. Such projections are
integrations in the spatial domain (in our case angular domain),
but due to the presence of an exponential term, ${\rm{e}}^{-Q}$,
these integrations can not be performed as desirable. Then, we
have engendered two schemes to overcome this difficulty (cf. Eqs.
(\ref{eq11}) and (\ref{eq12})), where the decomposition of the
exponential with respect to a series of Chebyshev functions using
the collocation method appeared to be quite satisfactory as
demonstrated by several numerical experiments. Accordingly, even
for a small truncation order, say $N=7$ (and $N+1$ collocation
points) to decompose ${\rm{e}}^{-Q}$, the error associated to the
constraint and residual equation are acceptable. The increase of
truncation error to $N=11$ - still a modest truncation order -
provides a much better accuracy and fast convergence.

Finally, the combination of Galerkin and collocation methods can
be very useful in studying other interesting nonlinear problems in
Cosmology and Gravitation.

The authors acknowledge the financial support of the Brazilian
agencies CNPq and CAPES.

\end{document}